# Development, validation and in-class evaluation of a sequence of clicker questions on Larmor precession of spin in quantum mechanics

Paul Justice, Emily Marshman, and Chandralekha Singh

*Department of Physics and Astronomy, University of Pittsburgh, Pittsburgh, PA, 15260*

Engaging students with well-designed clicker questions is one of the commonly used research-based instructional strategy in physics courses partly because it has a relatively low barrier to implementation. Moreover, validated robust sequences of clicker questions are likely to provide better scaffolding support and guidance to help students build a good knowledge structure of physics than an individual clicker question on a particular topic. Here we discuss the development, validation and in-class implementation of a clicker question sequence (CQS) for helping advanced undergraduate students learn about Larmor precession of spin, which takes advantage of the learning goals and inquiry-based guided learning sequences in a previously validated Quantum Interactive Learning Tutorial (QuILT). The in-class evaluation of the CQS using peer instruction is discussed by comparing upper-level undergraduate students' performance after traditional lecture-based instruction and after engaging with the CQS.

## I. INTRODUCTION AND BACKGROUND

Clicker questions (also known as concept tests) are conceptual multiple-choice questions typically administered in the classroom to engage students in the learning process and obtain feedback about their learning via a live feedback system called clickers [1-13]. Integration of peer interaction with lectures via clicker questions has been popularized in the physics community by Mazur [2]. In Mazur's approach, the instructor poses conceptual, multiple-choice clicker questions to students which are integrated throughout the lecture. Students first answer each clicker question individually, which requires them to take a stance regarding their thoughts about the concept(s) involved. Students then discuss their answers to the questions with their peers and learn by articulating their thought processes and assimilating their thoughts with those of the peers. Then after the peer discussion, they answer the question again using clickers followed by a general class discussion about those concepts in which both students and the instructor participate. The feedback that the instructor obtains is also valuable because the instructor has an estimate of the prevalence of common student difficulties and the fraction of the class that has understood the concepts and can apply them in the context in which the clicker questions are posed. The use of clickers keeps students alert during lectures and helps them monitor their learning. Clicker questions can be used in the classroom in different situations, e.g., they can be interspersed within lectures to evaluate student learning in each segment of a class focusing on a concept, at the end of a class, or to review materials from previous classes at the beginning of a class.

While clicker questions for introductory [2] and upper-level physics such as quantum mechanics [14] have been developed, there have been very few documented efforts [15-17] toward a systematic development and validation of clicker question sequences (CQSs), e.g., question sequences on a given concept that can be used in a few class periods when students learn the concepts and that build on each other effectively and strive to help students organize, extend and repair their knowledge structure pertaining to the topic.

Here we discuss the development, validation and in-class implementation of a CQS to help students develop conceptual understanding of the Larmor precession of spin in quantum mechanics (QM) that was developed for students in upper-level undergraduate QM courses taken by physics juniors and seniors. The CQS was developed by taking advantage of the learning goals and inquiry-based guided learning sequences in a research-validated Quantum Interactive Learning Tutorial (QuILT) on this topic [18] as well as by refining, fine-tuning and adding to the existing clicker questions from our group which have been individually validated previously [19-20]. The CQS can be used in class either separately from the QuILT or synergistically with the corresponding QuILT [18] if students engage with the QuILT after the CQS as another opportunity to reinforce the concepts learned.

## II. LEARNING GOALS AND METHODOLOGY

The learning goals and inquiry-based learning sequences in the QuILT, which guided the development and sequencing of the CQS questions, were developed using extensive research on student difficulties with these concepts as a guide and cognitive task analysis from an expert perspective.

### A. Learning goals

The first learning goal of the CQS (consistent with the QuILT) is to help students be able to unpack the consequence of Ehrenfest's theorem that there is no time dependence to the expectation value of any observable whose corresponding Hermitian operator commutes with the Hamiltonian regardless of the state of the quantum system. This is highlighted throughout the CQS by students considering the expectation value of $\hat{S}_z$ and realizing that it is always time independent regardless of the quantum state for the Hamiltonian $\hat{H} = -\gamma B_0 \hat{S}_z$ since $\hat{S}_z$ commutes with the Hamiltonian. The second learning goal is for students to learn another application of Ehrenfest's theorem in that the expectation value of any observable (which does not have explicit time-dependence) is not dependent on time when the initial state is a stationary state. In particular, if the system is in a stationary state (i.e., an eigenstate of the Hamiltonian) the expectation values of all observables are time independent, rather than just those observables whose corresponding operators commute with the Hamiltonian. Throughout the CQS, stationary states are eigenstates of $\hat{S}_z$, challenging students to realize that these are eigenstates of the Hamiltonian so the expectation values, e.g., of $\hat{S}_X$ and $\hat{S}_Y$ are also time-independent if the initial state is an eigenstate of $\hat{S}_z$. Finally, the third learning goal of the CQS is for students to be able to distinguish between stationary states and eigenstates of Hermitian operators that do not commute with the Hamiltonian (e.g., those corresponding to observables other than energy). For the Hamiltonian $\hat{H} = -\gamma B_0 \hat{S}_z$, students should learn that an eigenstate of either $\hat{S}_X$ and $\hat{S}_Y$ is not a stationary state, unlike a system in an eigenstate of $\hat{S}_z$.

### B. Development and validation

Based upon the learning goals delineated in the QuILT, questions in the Larmor precession of spin CQS were developed or adapted from prior validated clicker questions and sequenced to balance difficulties, avoid change of both concept and context between consecutive questions as appropriate in order to avoid a cognitive overload, and include a mix of abstract and concrete questions to help students develop a good grasp of relevant concepts. The validation was an iterative process. Individual questions were tweaked to address learning goals and known student difficulties over several years of in-class use, and sequencing was validated via in-class use and faculty feedback.

After the initial development of the Larmor precession of spin CQS using the learning goals and inquiry-based guided learning sequences in the QuILT and some existing individually validated clicker questions, we iterated the CQS with three physics faculty members who provided valuable feedback on fine-tuning and refining both the CQS as a whole and some new questions that were developed and adapted with existing ones to ensure that the questions were unambiguously worded and build on each other based upon the learning goals. In addition to interviews used in the validation of the QuILT, we also conducted individual think-aloud interviews with four advanced students who had learned these concepts via traditional lecture-based instruction in relevant concepts to confirm that they interpreted the CQS questions as intended and the sequencing of the questions provided appropriate scaffolding support to students.

The final version of the Larmor precession of spin CQS has 6 questions, which can be integrated with lectures in which these relevant concepts are covered in a variety of ways based upon the instructor's preferences. In particular, they can be interspersed with lecture or posed together depending, e.g., upon whether they are integrated with lectures similar to Mazur's approach, used at the end of each class or used to review concepts after students have learned via lectures everything related to Larmor precession of spin that the instructor wanted to teach.

The first two questions in the CQS, CQ1 and CQ2, begin by addressing the time-development of a state that is initially an energy eigenstate or not initially an energy eigenstate. This calls on students' prior knowledge about the time development of a state before addressing general characteristics of the time dependence of an expectation value of $\vec{\hat{S}}$ in CQ3 and CQ4. CQ5 addresses the time dependence of expectation value for different components of the spin for a state that is not an eigenstate of the Hamiltonian, but rather an eigenstate of the x-component of the spin angular momentum, $\hat{S}_X$. The sequence then concludes by contrasting CQ5 with a similar question CQ6 which is for a system initially in an energy eigenstate.

The six questions in the CQS are as follows:

**(CQ1)** An electron in a magnetic field $\vec{B} = B_0\hat{z}$ is initially in a spin state $|\chi(0)\rangle = |\uparrow\rangle_z$. Which of the following equations correctly represents the state $|\chi(t)\rangle$ of the electron after time t? The Hamiltonian operator is $\hat{H} = -\gamma B_0 \hat{S}_z$.
 a) $|\chi(t)\rangle = |\uparrow\rangle_z$
 **b) $|\chi(t)\rangle = e^{i\gamma B_0 t/2}|\uparrow\rangle_z$**
 c) $|\chi(t)\rangle = e^{i\gamma B_0 t/2}|\uparrow\rangle_z + e^{-i\gamma B_0 t/2}|\downarrow\rangle_z$
 d) $|\chi(t)\rangle = ae^{i\gamma B_0 t/2}|\uparrow\rangle_z + be^{i\gamma B_0 t/2}|\downarrow\rangle_z$
 e) None of the above

**(CQ2)** An electron in a magnetic field $\vec{B} = B_0\hat{z}$ is initially in a spin state $|\chi(0)\rangle = a|\uparrow\rangle_z + b|\downarrow\rangle_z$. Which of the following equations correctly represents the state $|\chi(t)\rangle$ of the electron after time t? The Hamiltonian operator is $\hat{H} = -\gamma B_0 \hat{S}_z$.
 a) $|\chi(t)\rangle = e^{i\gamma B_0 t/2}(a|\uparrow\rangle_z + b|\downarrow\rangle_z)$
 b) $|\chi(t)\rangle = e^{-i\gamma B_0 t/2}(a|\uparrow\rangle_z + b|\downarrow\rangle_z)$
 c) $|\chi(t)\rangle = e^{i\gamma B_0 t/2}((a+b)|\uparrow\rangle_z + (a-b)|\downarrow\rangle_z)$
 **d) $|\chi(t)\rangle = ae^{i\gamma B_0 t/2}|\uparrow\rangle_z + be^{-i\gamma B_0 t/2}|\downarrow\rangle_z$**
 e) None of the above

**(CQ3)** Choose all of the following statements that are true about the expectation value $\langle\vec{\hat{S}}\rangle$ for an electron in a magnetic field $\vec{B} = B_0\hat{z}$ in the state $|\chi(t)\rangle$ when the initial state is **NOT** $|\uparrow\rangle_z$ or $|\downarrow\rangle_z$. The Hamiltonian operator is $\hat{H} = -\gamma B_0 \hat{S}_z$.
 I. The z-component of $\langle\vec{\hat{S}}\rangle$, i.e. $\langle\hat{S}_Z\rangle$, is time-independent.
 II. The x- and y-components of $\langle\vec{\hat{S}}\rangle$ change with time. When the magnitude of $\langle\hat{S}_X\rangle$ is a maximum, the magnitude of $\langle\hat{S}_Y\rangle$ is a minimum, and vice versa.
 III. The magnitudes of the maximum values of $\langle\hat{S}_X\rangle$ and $\langle\hat{S}_Y\rangle$ are the same.
 a) I only  b) I and II only
 c) I and III only  d) II and III only
 **e) All of the above**

**(CQ4)** Choose all of the following statements that are true about the expectation value $\langle\vec{\hat{S}}\rangle$ for an electron in a magnetic field $\vec{B} = B_0\hat{z}$ in the state $|\chi(t)\rangle$ when the initial state is **NOT** $|\uparrow\rangle_z$ or $|\downarrow\rangle_z$. The Hamiltonian operator is $\hat{H} = -\gamma B_0 \hat{S}_z$.
 I. The vector $\langle\vec{\hat{S}}\rangle$ can be thought to be precessing about the z-axis at a non-zero angle.
 II. The vector $\langle\vec{\hat{S}}\rangle$ can be thought to be precessing about the z-axis at a frequency $\omega = \gamma B_0$.
 III. All three components of vector $\langle\vec{\hat{S}}\rangle$ change as it precesses about the z-axis.
 a) I only  **b) I and II only**
 c) I and III only  d) II and III only
 e) All of the above

**(CQ5)** Suppose an electron in a magnetic field $\vec{B} = B_0\hat{z}$ is initially in an eigenstate of the x-component of spin angular momentum operator, i.e. $\hat{S}_X$. The Hamiltonian operator is $\hat{H} = -\gamma B_0 \hat{S}_z$. Choose all of the following statements that are correct.
 I. The expectation value $\langle\hat{S}_X\rangle$ depends on time.
 II. The expectation value $\langle\hat{S}_Y\rangle$ depends on time.
 III. The expectation value $\langle\hat{S}_Z\rangle$ depends on time.
 a) I only  b) III only
 **c) I and II only**  d) II and III only
 e) All of the above

**(CQ6)** Suppose an electron in a magnetic field $\vec{B} = B_0\hat{z}$ is initially in an eigenstate of the z-component of spin angular momentum operator, i.e. $\hat{S}_Z$. The Hamiltonian operator is $\hat{H} = -\gamma B_0 \hat{S}_z$. Choose all of the following statements that are correct.
 I. The expectation value $\langle\hat{S}_X\rangle$ depends on time.
 II. The expectation value $\langle\hat{S}_Y\rangle$ depends on time.

*III. The expectation value $\langle \hat{S}_z \rangle$ depends on time.*
   *a) I only*     *b) III only*
   *c) I and II only*  ***d) None of the above***
   *e) All of the above*

### C. In-class implementation

The final version of the CQS on the Larmor precession of spin was implemented with peer discussion [2-4] in two upper-level undergraduate QM classes at a large research university (Pitt) after traditional lecture-based instruction in relevant concepts in two consecutive years (primarily white students). Prior to the implementation of the CQS in both classes with peer interaction, students took a pretest after traditional lecture-based instruction. The pre/posttests were developed and validated by Brown and Singh [18] to measure comprehension of the concepts related to the time-dependence of expectation values of observables in the context of the Larmor precession of spin. The CQS was implemented right after the pretest in one class period with peer interaction. The posttest was administered during the following week to measure the impact of the CQS.

On the pretest and posttest, students were given that the Hamiltonian of the system is $\hat{H} = -\gamma B_0 \hat{S}_z$ with questions 1-3 being analogous but different and questions 4-6 being identical. In particular, an electron is initially in an eigenstate of $\hat{S}_x$ ($\hat{S}_Y$ on the posttest) in questions 1-3, and students are asked if the expectation value of $\hat{S}_x$, $\hat{S}_Y$, and $\hat{S}_Z$ respectively depend on time. Students are also expected to explain their reasoning. These questions primarily focus on the first and third learning goals. Question 4 presents the following conversation between two students about an electron initially in an eigenstate of $\hat{S}_x$ ($\hat{S}_Y$ on the posttest) and asks with whom they agree. This question addresses the first and third learning goals.

**Andy**: *The electron will NOT be in an eigenstate of $\hat{S}_x$ forever because the state will evolve in time.*
**Caroline**: *I disagree. If a system is in an eigenstate of an operator corresponding to a physical observable, it stays in that state forever unless a perturbation is applied.*

Questions 5 asks students if the expectation value of $\hat{S}_Y$ is time dependent if the initial state of the system is an eigenstate of $\hat{S}_Z$ (i.e., an eigenstate of the Hamiltonian or stationary state). Then, question 6 asks if there is precession around the z-axis for an electron initially in an eigenstate of $\hat{S}_Y$, and if so, to give an example of a situation in which there would be no precession. Both of these questions deal with the second and third learning goals. All questions ask students to justify their answers. Partial credit was awarded to students who answered correctly, but with no or inadequate justification, consistent with the agreed upon rubric. Interrater reliability between the two researchers who graded all pre/posttests was above 95%.

### III. IN-CLASS IMPLEMENTATION RESULTS

Tables 1-3 compare average pre/posttest performances of students on each question in the upper-level QM course from the same large research university in two different years after traditional lecture-based instruction (pretest) and on the posttest after students had engaged with the CQS with peer instruction on the Larmor precession of spin (Table 1-2 are for the two classes separately and Table 3 is for the two classes combined). The normalized gain (or gain) is calculated as $g = (post\% - pre\%)/(100\% - pre\%)$ [21]. Similarly, the effect size is calculated for all questions in all tables. Effect size is calculated as Cohen's $d = (\mu_{post} - \mu_{pre})/\sigma_{pooled}$ where $\mu_i$ is the mean of group $i$ and the pooled standard deviation is $\sigma_{pooled} = \sqrt{(\sigma_{pre}^2 + \sigma_{post}^2)/2}$ [22].

TABLE 1. Comparison of mean pre/posttest scores on each question, normalized gains (g) and effect sizes Cohen's d (d) for upper-level undergraduate QM students in class A who engaged with the CQS on Larmor precession of spin concepts (N=17).

| Q# | Pretest Mean | Posttest Mean | g | d |
|---|---|---|---|---|
| 1 | 22% | 75% | 0.68 | 0.69 |
| 2 | 47% | 84% | 0.71 | 0.54 |
| 3 | 19% | 72% | 0.65 | 0.66 |
| 4 | 31% | 81% | 0.73 | 0.62 |
| 5 | 13% | 47% | 0.39 | 0.42 |
| 6 | 34% | 75% | 0.62 | 0.50 |

TABLE 2. Comparison of mean pre/posttest scores on each question, normalized gains (g) and effect sizes Cohen's d (d) for upper-level undergraduate QM students in class B who engaged with the CQS on Larmor precession of spin concepts (N=39).

| Q# | Pretest Mean | Posttest Mean | g | d |
|---|---|---|---|---|
| 1 | 50% | 92% | 0.85 | 0.56 |
| 2 | 67% | 94% | 0.82 | 0.48 |
| 3 | 52% | 95% | 0.91 | 0.61 |
| 4 | 41% | 79% | 0.64 | 0.43 |
| 5 | 38% | 82% | 0.71 | 0.51 |
| 6 | 56% | 85% | 0.66 | 0.35 |

TABLE 3. Comparison of mean pre/posttest scores on each question, normalized gains (g) and effect sizes Cohen's d (d) for upper-level undergraduate QM students in both class A and class B combined who engaged with the CQS on Larmor precession of spin concepts (N=56).

| Q# | Pretest Mean | Posttest Mean | g | d |
|---|---|---|---|---|
| 1 | 41% | 87% | 0.78 | 0.59 |
| 2 | 60% | 91% | 0.77 | 0.49 |
| 3 | 41% | 88% | 0.79 | 0.59 |
| 4 | 38% | 80% | 0.67 | 0.49 |
| 5 | 30% | 70% | 0.58 | 0.46 |
| 6 | 49% | 82% | 0.64 | 0.40 |

All three tables show moderate effect sizes from pre to posttest on each of the questions, with all effect sizes above 0.3, and some even nearing 0.7. Additionally, all normalized gains exceed 0.3, with most falling in the range of 0.6-0.9. Despite varied pretest scores for the two classes on different questions, posttest scores for both classes on most questions demonstrate that the CQS is effective in addressing the learning goals.

Although the two groups of upper-level physics majors in the QM course from the same university in two consecutive years are different, the difference in pretest scores between two classes may also be a reflection of the difference between the effectiveness of traditional instruction of the two instructors. On a positive note, the posttest scores for both groups are relatively robust. We note that overall the CQS implementation was consistent between the two years, and both instructors provided the same class participation credit for clicker questions and low-stakes testing credit for students to take the pre-/posttests. However, no constraints were placed on instructors' teaching of the topic in class prior to the implementation of the CQS, as the CQS in this implementation is meant to act primarily as a "second coat" to reinforce learning. Moreover, possible differences between the instructors may include, but are certainly not limited to, the differences in pedagogy and the time spent in lecture on the topic.

Since the researchers did not have control over traditional instruction, we focus on the posttest scores after the CQS. Tables 1 and 2 show that the difference in pretest scores between the two classes is also followed by a corresponding difference in posttest scores after the CQS. These differences in posttest scores may indicate that the prior knowledge of the material (or the first coat) does affect how well students learn from the CQS implementation with peer instruction. In particular, Class B, which exhibited higher pretest scores than Class A on all test questions, also exhibited higher posttest scores on five of the six test questions, and exhibited comparable posttest scores on question 4. From this comparison, we conclude that the CQS did not eliminate the performance gap in pretest resulting from differences in knowledge after traditional instruction in the two classes and before students engaged with the CQS. It is possible that certain threshold knowledge may be prerequisite for optimal learning from the CQS particularly because peer instruction was involved and students can meaningfully communicate and learn from each other only if together they have certain threshold knowledge.

We note that those higher pretest scores of Class B in Table 2 do not exceed the posttest scores of Class A in Table 1. While the highest average score for Class B on the pretest is still below 70%, five of the six posttest scores for Class A exceed 70%. Thus, regardless of the effectiveness of traditional lecture-based instruction for a given instructor, students still gained from the CQS. Moreover, as shown in Table 3, overall (averaging over the two classes), the CQS was effective in addressing its learning goals. On average, student performance on test questions range from 30-60% on the pretest, showing that there is much room for improvement after traditional instruction. After the implementation of the CQS, average scores on test questions exceeded 70% on the posttest (roughly 90% on Q# 1-3).

Difficulty with question 5 was the most common among students, with Class A averaging below 50% on the question even after the CQS implementation. This suggests that there is still room for improving the CQS when dealing with the second learning goal in order for students to understand the special role of the stationary states in different contexts. This was a learning goal addressed by both questions 5 and 6 on the pretest and posttest, but the difference in performance suggests that students with less prior knowledge (Class A) failed to perform on question 5, even though they averaged above 75% on question 6, which provides more scaffolding. In order to address this difficulty, the CQS may be improved by adding another question later in the CQS that more directly addresses this second learning goal. This question could provide an opportunity to wean students from the scaffolding provided in prior questions related to this learning goal and may allow for more effective learning on future implementations of this CQS.

## IV. SUMMARY

Clicker questions are relatively easy to implement in the classroom alongside more traditional lecture-based instruction. We developed and validated a clicker question sequence related to the time-dependence of expectation values in the context of Larmor precession of spin that continually builds on students' knowledge as they engage with different questions in the CQS after traditional instruction in relevant concepts. Throughout the development and validation which was an iterative process, many students and instructors provided feedback several times. The in-class implementation of the CQS in two upper-level quantum mechanics classes shows moderate effect sizes for gains in students' performance from the pretest to posttest suggesting this CQS is effective in helping students learn these concepts. The differences in the pretest scores in the two classes could be due to the differences in the students and the instructors and the effectiveness of their traditional instruction over which the researchers did not have any control. However, the posttest scores on all questions in both classes were reasonable, suggesting that the CQS is effective regardless of efficacy of an instructor's traditional lecture-based instruction. Moreover, comparison of the pre/posttest scores on each question for the two classes may shed some light on the role of prior knowledge upon which students can build as they engage with the CQS questions. In particular, the average posttest scores were generally higher on each question for the class which had a higher pretest scores. This issue will be investigated further in future implementation.

## ACKNOWLEDGMENTS

We thank the NSF for award PHY-1806691.

[1] C. Henderson and M. Dancy, Barriers to the use of research-based instructional strategies: The influence of both individual and situational characteristics, Phys. Rev. PER **3**, 020102 (2007).
[2] E. Mazur, *Peer Instruction: A User Manual*, Prentice Hall, Upper Saddle River N.J., (1997).
[3] C. Crouch and E. Mazur, Peer Instruction: Ten years of experience and results, Am. J. Phys. **69**, 970 (2001).
[4] A. Fagen, C. Crouch and E. Mazur, Peer Instruction: Results from a range of classrooms, The Phys. Teach. **40**, 206 (2002).
[5] D. Meltzer and K. Manivannan, Transforming the lecture-hall environment: The fully interactive physics lecture, Am. J. Phys. **70**, 639 (2002).
[6] E. Judson, Learning from past and present: Electronic response systems in college lecture halls, J. Comp. Math. Sci. Teach. **21**, 167 (2002).
[7] M. James, The effect of grading incentive on student discourse in Peer Instruction, Am. J. Phys. **74**, 689 (2006).
[8] S. Willoughby and E. Gustafson, Technology talks: Clickers and grading incentive in the large lecture hall, Am. J. Phys. **77,** 180 (2009).
[9] N. Lasry, J. Watkins, and E. Mazur, Peer instruction: From Harvard to the two-year college, Am. J. Phys. **76**, 1066 (2008).
[10] M. James and S. Willoughby, Listening to student conversations during clicker questions: What you have not heard might surprise you! Am. J. Phys. **79**, 123 (2011).
[11] G. Novak G, E. Patterson, A. Gavrin, and W. Christian, *Just-in-Time Teaching: Blending Active Learning with Web Technology*, Upper Saddle River, NJ: Prentice Hall (1999).
[12] N. Karim, A. Maries and C. Singh, Impact of evidence-based flipped or active-engagement non-flipped courses on student performance in introductory physics, Can. J. Phys. **96**, 411 (2018).
[13] N. Karim, A. Maries and C. Singh, Do evidence-based active-engagement courses reduce the gender gap in introductory physics?, Eur. J. Phys. **39**, 025701 (2018).
[14] R. Sayer, E. Marshman, and C. Singh, A case study evaluating Just-in-Time Teaching and Peer Instruction using clickers in a quantum mechanics course, Phys. Rev. PER **12**, 020133 (2016).
[15] L. Ding, N. Reay, A. Lee and L. Bao, Are we asking the right questions? Validating clicker question sequences by student interviews, Am. J. Phys. **77,** 643 (2009).
[16] P. Justice, E. Marshman and C. Singh, Improving student understanding of quantum mechanics underlying the Stern-Gerlach experiment using a research-validated multiple-choice question sequence, Eur. J. Phys. **40**, 055702 (2019).
[17] P. Justice, E. Marshman, and C. Singh, Development and validation of a sequence of clicker questions for helping students learn addition of angular momentum in quantum mechanics, Physics Education Research Conference Proc., (2018) https://doi.org/10.1119/perc.2018.pr.Justice
[18] B. Brown and C. Singh, Development and evaluation of a quantum interactive learning tutorial on Larmor precession of spin, Physics Education Research Conference Proc. (2015) https://doi.org/10.1119/perc.2014.pr.008.
[19] C. Singh and G. Zhu, Improving students' understanding of quantum mechanics by using peer instruction tools, Physics Education Research Conference Proc., AIP Conf. Proc., Mellville, New York **1413**, 77-80 (2012). https://doi.org/10.1063/1.3679998
[20] G. Zhu and C. Singh, Peer instruction for quantum mechanics, APS Forum on Education Newsletter, 8-10, Fall (2009) https://www.aps.org/units/fed/newsletters/spring2004/7pitt.html
[21] R. Hake, Interactive engagement versus traditional methods: A six-thousand student survey of mechanics test data for introductory physics courses, Am. J. Phys. **66**, 64 (1998).
[22] J. Cohen, *Statistical Power Analysis for the Behavioral Sciences*, Lawrence Erlbaum Associates, London (1988).